  \documentclass[apj,numberedappendix]{emulateapj}                    %
  \usepackage{apjfonts}                                               %
 
\usepackage{epsfig}

\newcommand{\be}{\begin{equation}}
\newcommand{\ee}{\end{equation}}
\newcommand{\ba}{\begin{eqnarray}}
\newcommand{\ea}{\end{eqnarray}}

\newcommand{\saxj}{\mbox{SAX~J1808.4-3658}}
\newcommand{\herx}{\mbox{Her~X-1}}

\newcommand{\ud}{\mathrm{d}}
\newcommand{\rb}{\bar{r}}
\newcommand{\textdegree}{\ensuremath{^\circ}}

\slugcomment{Accepted by The Astrophysical Journal.}

\shorttitle{Pulse Shapes from Rapidly-Rotating Neutron Stars}

\begin{document}

\title{Pulse Shapes from Rapidly-Rotating Neutron Stars: Equatorial Photon Orbits}
\author{Coire Cadeau\altaffilmark{1}, Denis A. Leahy\altaffilmark{2} and Sharon M. Morsink\altaffilmark{1}}
\email{ccadeau@phys.ualberta.ca, leahy@iras.ucalgary.ca, morsink@phys.ualberta.ca}
\altaffiltext{1}{Theoretical Physics Institute, Department of Physics,
University of Alberta, Edmonton, AB, T6G~2J1, Canada}
\altaffiltext{2}{Department of Physics and Astronomy, University of Calgary,
Calgary AB, T2N~1N4, Canada}

\begin{abstract}
We demonstrate that fitted values of stellar radius obtained by
fitting theoretical light curves to observations of millisecond period
X-ray pulsars can significantly depend on the method used to calculate
the light curves.  The worst-case errors in the fitted radius are
evaluated by restricting ourselves to the case of light emitted and
received in the equatorial plane of a rapidly-rotating neutron star.
First, using an approximate flux which is adapted to the
one-dimensional nature of such an emission region, we show how pulse
shapes can be constructed using an exact spacetime metric and fully
accounting for time-delay effects. We compare this to a method which
approximates the exterior spacetime of the star by the Schwarzschild
metric, inserts special relativistic effects by hand, and neglects
time-delay effects.  By comparing these methods, we show that there
are significant differences in these methods for some applications,
for example pulse timing and constraining the stellar radius.  In the
case of constraining the stellar radius, we show that fitting the
approximate pulse shapes to the full calculation yields errors in the
fitted radius of as much as about $\pm 10\%$, depending on the
rotation rate and size of the star as well as the details describing
the emitting region.  However, not all applications of pulse shape
calculations suffer from significant errors: we also show that the
calculation of the soft-hard phase lag for a 1~keV blackbody does not
strongly depend on the method used for calculating the pulse shapes.
\end{abstract}


\section{Introduction}
Observations of pulsed light emitted from the surface of a neutron
star have the potential to constrain the star's mass, radius and
equation of state. By modeling the physics of the emission region
(such as the emissivity, shape and size) and tracing the paths of
photons traveling through the relativistic gravitational field of the
spinning neutron star to the observer, it is possible to create model
pulse shapes which can be compared to the observed pulse shapes,
allowing fits to the star's macroscopic parameters to be made. The
main problem in such a program is disentangling the effects arising
from the neutron star's gravitational field and the effects coming
from the assumptions about the spectrum and geometry of the emission
region. A further complication arises if the neutron star is rotating
rapidly, since the effects of rotation can significantly alter the
geodesic paths traveled by the photons from the simple Schwarzschild
geodesics.

Radio pulsars have most of their light emitted far from the neutron
star, thus they are unlikely to be useful for constraining the neutron
star's properties.  However, accreting X-ray pulsars, whose light is
emitted from an accretion column close to the surface, provide a good
probe of the strong gravitational field near the neutron
star. Modeling of observed pulse shapes was first carried out without
including gravitational effects using polar cap models of increasing
complexity \citep{wan81,lea90,lea91}. The later models were motivated
by radiative transfer calculations of the emissivity by caps and
cylindrical columns by \citet{mes85}. Gravitational light-bending was
early on realized to be important \citep{PFC83} and was included in
model calculations of caps and columns by \citet{mes88} and in fits 
of cap models to observed pulse profiles by \citet{lea95}.  Accretion
column models with light-bending were computed by \citet{kra01},
\citet{lea03} and \citet{kra03}.  Analysis of the occultation sequence
of the X-ray pulsar \herx\ by \citet{sco00} gave the unique
observational result that the pulse shape must be due to a pencil beam
from the near pole and a gravitationally-focused fan beam from the far
pole. This resulted in a quantitative model for the pulse shape of
\herx\ \citep{lea04} and a mass-to-radius constraint for the neutron
star \citep{lea04b}.

For the X-ray pulsars that have longer than millisecond periods, the
rotation rates are so slow (the fastest is 69 ms, most are $>1$ s)
that it is not necessary to introduce any rotational corrections at
all.  The first millisecond X-ray pulsar in a low-mass X-ray binary,
\saxj, was recently discovered by \citet{WvdK98} and
spins at a frequency of 400 Hz \citep{CM98}; this is fast enough that
the rotational speeds at the equator must be relativistic.  In
addition, X-ray bursts from a number of rapidly-rotating neutron stars
in low-mass X-ray binaries have been observed to have phase lags
between different energy bands \citep{CMT98}.  These phase lags may 
have their origin in relativistic effects.

A standard approximation scheme for ray-tracing near rotating neutron
stars has been to treat the propagation of photons as though the
gravitational field were static, so that the Schwarzschild metric and
the formalism presented by \citet{PFC83} can be used. The effects of
rotation are brought into the problem by introducing special
relativistic Doppler boosts as though the star had no gravitational
field. We call this approximation ``Schwarzschild + Doppler''
(S+D). Recently the S+D approximation has been used by \citet{PG03} to
fit the pulse profile of \saxj\ and to place constraints on
the neutron star's value of $M/R$.

One would expect that the S+D approximation is reasonable for slow
rotation or for fast rotation with emission near the spin poles of the
star. However, if emission occurs from a region close to the equator
of a rapidly-rotating neutron star, it is possible that this
approximation scheme could break down. One of the main goals of this
paper is to evaluate the accuracy of this approximation by comparing
against an exact treatment of photon propagation in the equatorial
plane, where we expect the largest errors to occur.  To accomplish
this, we consider a simple model where the emitting region is a
one-dimensional curve that emits photons only into the equatorial
plane.  By considering a variety of calculation methods and types of
emission for this one-dimensional test case, we can quantify the
worst-case errors that can arise in applications using approximate
versions of the pulse shape calculation.

The effects of relativistic rotation on pulse shapes have been
considered in a number of different treatments.  The timing of pulses
from millisecond pulsars in the polar cap model where emission comes
from a radially-extended region near the pulsar's surface was
investigated using the metric for a slowly-rotating neutron star
\citep{KD86}, and the Kerr black hole metric \citep{Kapoor91}.
The lag of low energy photons during X-ray bursts was modeled by
including special relativistic Doppler effects but ignoring all
gravitational effects \citep{Ford99}. This work was later extended
using the S+D approximation in order to model the energy-dependent
delays in \saxj\ \citep{Ford00}.  The oscillation amplitudes and
energy phase lags during X-ray bursts on rapidly-rotating neutron
stars have been investigated using the S+D approximation without
time-delays \citep{ML98,WML01} and with time-delays \citep{MOC02} and
also using the Schwarzschild metric with neither Doppler effects nor
time delays \citep{NSS02}. Pulse shapes have also been computed by
using the rotating Kerr black hole metric to approximate the spacetime
of a rotating neutron star \citep{CS89,BRR00}.

In this paper we are evaluating the accuracy of the S+D approximation
against the exact calculation of a rapidly-rotating neutron star. In
order to evaluate the largest extent of the errors, we are restricting
our calculations to photons emitted from the equator which travel in
the equatorial plane to the observer. In Section~\ref{s:rapid} we
outline the details of the calculation of pulse shapes for equatorial
photons emitted by a rapidly-rotating neutron star. In
Section~\ref{s:approx} we briefly review the standard S+D
approximation and compare its predictions against those of
Section~\ref{s:rapid} for a simple type of emission. In order to
evaluate the importance of these differences, we compute the effects
of different emission models on pulse shapes in
Section~\ref{s:emission}.  The effects of the approximation on the
fitted neutron star radius are shown in Section~\ref{s:fit}.

\section{Pulse Shapes for Rapidly-Rotating Neutron Stars}
\label{s:rapid}
A two-dimensional area of a star emits light with frequency $\nu_e$,
intensity $I_{\nu_e}$ in a direction which makes an angle of
$\alpha_e$ with respect to the normal to the surface. An observer far
from the star sees that the area subtends a solid angle $\ud\Omega_o$
and the photons have an observed frequency $\nu_o$.  Defining the
redshift $z$ through the relation $1+z = \nu_e/\nu_o$, and making use
of the conservation of photon number density in phase space, the
observed flux is
\be
F_{\nu_o} = I_{\nu_o}\ud\Omega_o 
		= I_{\nu_e} \left(\frac{ 1}{1+z}\right)^3\ud\Omega_o.
\ee
In order to construct a light curve if the area produced an
instantaneous flash of light, we must take into account the fact that
the emitting surface is moving and that photons from different parts
of the surface arrive at the detector at different times. In this
section we describe how the observed angle subtended by the emitting
area and the times of arrival are computed, and light curves
constructed.

We are interested in modeling light propagation on the background of a
rapidly-rotating neutron star. Accurate models of rotating neutron
stars for tabulated equations of state can be computed numerically
using the public-domain code \texttt{rns}\footnote{The \texttt{rns}
code is available at \url{http://www.gravity.uwm.edu/rns/}}
\citep{SF95}. This computer code computes the metric functions
$\alpha$, $\gamma$, $\rho$, and $\omega$ appearing in the axisymmetric,
stationary metric
\be\label{eq:metric}
\ud s^{2} = -e^{\gamma + \rho} \ud t^{2} 
            + e^{\gamma - \rho} \bar{r}^{2} \sin^{2} \theta 
			(\ud \phi - \omega \ud t)^{2}
            + e^{2\alpha}(\ud \bar{r}^{2} + \bar{r}^{2} \ud\theta^{2}),
\ee
where the metric functions depend only on the coordinates $\theta$ and
$\bar{r}$. The metric function $\omega$ is the term responsible for
the frame-dragging effect and would vanish if the star weren't
rotating.  The coordinate $\bar{r}$ corresponds to the isotropic
Schwarzschild radial coordinate in the limit of zero rotation. The
standard radial coordinate that appears in the Schwarzschild metric,
$r$, is related to our coordinates by $r = \bar{r}
\exp{\frac12(\gamma-\rho)}$.  In the limit of zero rotation, the
following combinations of metric functions are:
\ba
\lim_{\Omega_\star\rightarrow0}  e^{\frac12(\gamma+\rho)} &=& \left( 1- \frac{2M}{r} \right)^{1/2}, 
\label{eq:lim1}\\
\lim_{\Omega_\star\rightarrow0}  \bar{r} e^{-\rho} &=& r \left( 1- \frac{2M}{r} \right)^{-1/2},\ \mathrm{and}
\label{eq:lim2}\\
\lim_{\Omega_\star\rightarrow0}  e^{\alpha-\frac12(\gamma+\rho)}\ud\bar{r} 
 &=& \left( 1- \frac{2M}{r} \right)^{-1}\ud r,
\label{eq:lim3}
\ea
where $\Omega_\star$ is the star's angular velocity, as measured by an observer at infinity.

\subsection{Selection of Stellar Models}
We have chosen four stellar models to illustrate the effects discussed
in this paper. We have chosen equations of state (EOS) A and L from the
\citet{AB77} catalogue which span a realistic range of stiffness.  EOS
A is one of the softest equations of state and EOS L is one of the
stiffest allowed by present observations. For each EOS, we have
computed two $1.4M_\odot$ models with spin frequencies 300 Hz and 600
Hz, spanning the range of observed spin frequencies seen during X-ray
bursts. The parameters describing these models are given in
Table~\ref{tab:modelparams}. We have also chosen to designate Model 4
(EOS L, 600 Hz) as our fiducial model against which all approximations
will be made.  We have selected this model since it is both the
fastest and largest model of the set, so we expect that any effects
due to relativistic velocities or long light-crossing times will be
maximized in this model.

\begin{deluxetable*}{cclllllll}
\tablecaption{Neutron Star Models with Mass $= 1.4 M_\odot$ \label{tab:modelparams}}
\tablehead{
\colhead{Model} &
\colhead{EOS} &
\colhead{$\Omega_{B}/2\pi$\tablenotemark{a} (Hz)} &
\colhead{$\Omega_\star/2\pi$ (Hz)} &
\colhead{$R$\tablenotemark{b} (km) } &
\colhead{$cJ/(G M^{2})$} &
\colhead{$GM/(c^{2}R)$} &
\colhead{$v/c$\tablenotemark{c}} &
\colhead{$\omega_{eq}/2\pi$\tablenotemark{d} (Hz)}
}
\startdata
1 & A    & 1387 & 300 &  9.62 & 0.109 & 0.21 & 0.08 & 50.2 \\*
2 &      &      & 600 &  9.78 & 0.223 & 0.21 & 0.16 & 98.3 \\[0.25em]
3 & L    & 742  & 300 & 15.11 & 0.234 & 0.14 & 0.11 & 27.9 \\*
4 &      &      & 600 & 16.38 & 0.508 & 0.13 & 0.24 & 48.9 \\
\enddata
\tablenotetext{a}{The break-up spin frequency for a star with the given mass and equation of state.}
\tablenotetext{b}{The equatorial Schwarzschild radius.}
\tablenotetext{c}{The speed of the neutron star at the equator measured by a static
	          observer at the surface.  Velocities are calculated with the full metric.}
\tablenotetext{d}{The frame-dragging term at the equator; this is the
		  angular velocity of a zero angular momentum particle at the
		  equator.}
\label{tab:1}
\end{deluxetable*}

We are interested in quantifying differences between the S+D
approximation and the exact rotation calculation. In order to do so,
we will refer to ``Schwarzschild Equivalent'' (SE) stars, which are
spherically-symmetric stars with the same mass and equatorial radius
as the models listed above. Note that these SE stars have a larger
radius than would be normally computed for a static star with the
given EOS and mass. The equatorial velocity appearing in the Doppler
shift factors used in the S+D approximation is
\be
v_{eq} = \Omega_\star \frac{R}{\sqrt{1-2M/R}}.
\label{eq:veq}
\ee
In the full relativistic calculation a static observer located at the
surface of the star measures the fluid at the equator to have the
velocity
\be
v = \frac{\Omega_\star \rb e^{-\rho}}{1 + \omega\rb^2 e^{-2\rho}(\Omega_\star - \omega)}.
\label{eq:veqgen}
\ee
In the limit of zero frame-dragging, $v$ reduces to $v_{eq}$ when
Equation~\ref{eq:lim2} is used. The values of the frame-dragging term
$\omega$ at the equator and the velocity $v$ are given in
Table~\ref{tab:modelparams}. Since the frame-dragging frequency is a
small fraction of the star's spin frequency, its neglect in the S+D
approximation should not be very important.

\subsection{Photon Paths in the Equatorial Plane}
\label{sec:null}
In this paper we restrict ourselves to photons emitted from the
equator in a direction in the equatorial plane. With this restriction,
the motion of a photon is specified once the initial location and
impact parameter are specified. The geodesic equations describing the
paths traced by outgoing photons are:
\begin{eqnarray}
\label{eq:nullt}   \left(\frac{\ud t}{\ud \lambda}\right)    & = & e^{-(\gamma + \rho)}(1 - \omega b), \\
\label{eq:nullphi} \left(\frac{\ud \phi}{\ud \lambda}\right) & = & \omega e^{-(\gamma + \rho)}(1 - \omega b) 
							+ \frac{b}{\bar{r}^{2}} e^{\rho - \gamma},\ \mathrm{and}\\
\label{eq:nullr}   \left(\frac{\ud \bar{r}}{\ud \lambda}\right) & = & e^{-\alpha-\frac12(\gamma+\rho)} 
			\left( (1-\omega b)^{2} 
				-\frac{b^{2}}{\bar{r}^{2}} e^{2\rho} \right)^{1/2},
\end{eqnarray}
where $b$ is the photon's impact parameter and $\lambda$ is an affine
parameter defined so that photon orbits are independent of energy.  We
are considering photons originating on the equatorial plane ($\theta =
\pi/2$) emitted parallel to the equatorial plane ($u^\theta = 0$
initially).  It is a straightforward calculation to show that such
photons must remain in the equatorial plane, i.e., $\ud \theta / \ud
\lambda = 0$.

\begin{figure*}
\epsscale{1.0}
\plotone{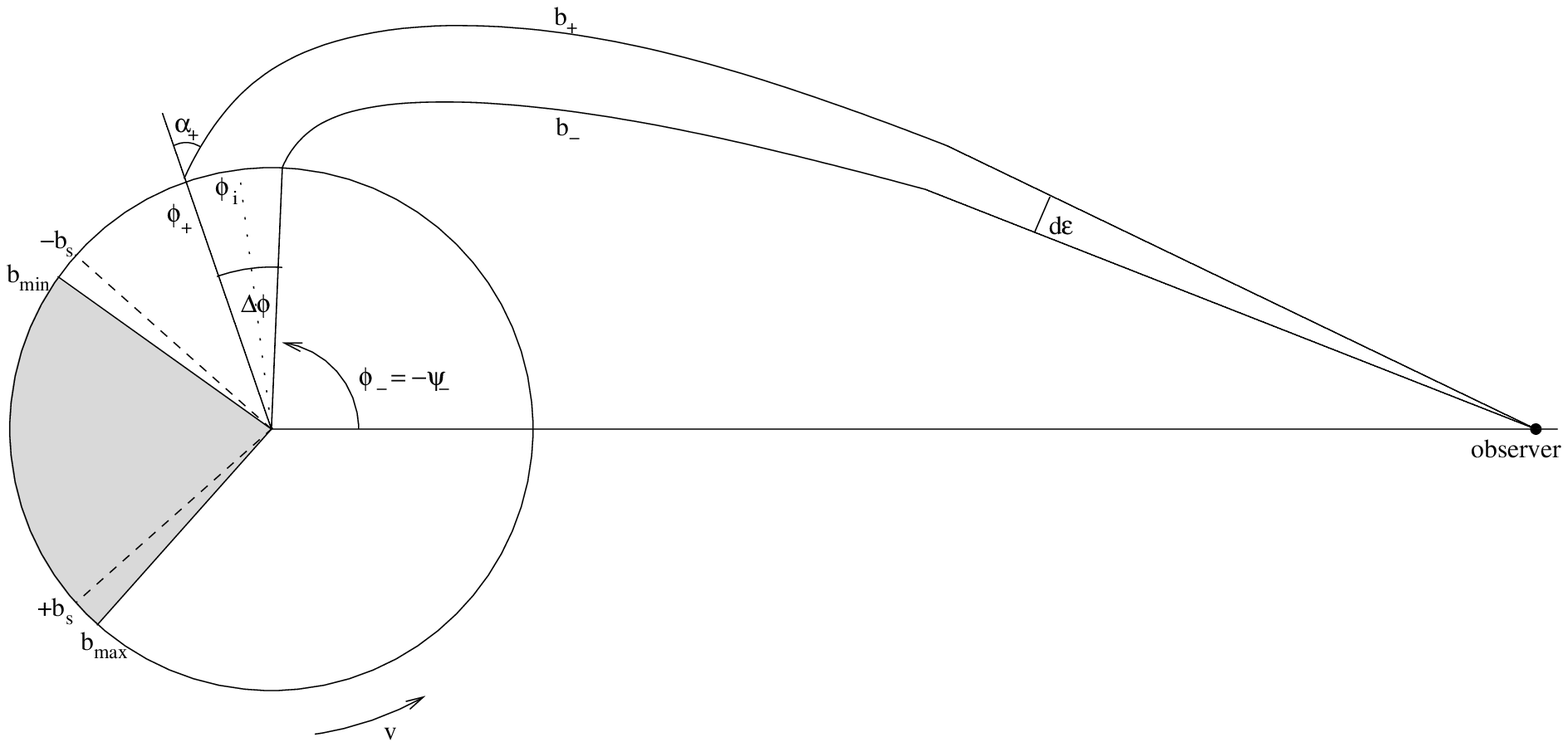}
\caption{Definition of angles.} 
\label{fig:angles}
\end{figure*}

Since the radial component of the four-velocity must be real,
the impact parameters must lie in the range $b_{\mathrm{min}} \le b
\le b_{\mathrm{max}}$, where the minimum and maximum impact parameters
are:
\begin{eqnarray}
b_{\mathrm{min}} & = & - \bar{r} e^{-\rho} \frac{1}{1 - \omega \bar{r}e^{-\rho}},\ \mathrm{and}\\
b_{\mathrm{max}} & = & \bar{r} e^{-\rho} \frac{1}{1 + \omega \bar{r}e^{-\rho}},
\end{eqnarray}
where the metric potentials are to be evaluated at the point at which
the null ray originates, e.g., the surface of the star. The
frame-dragging term is positive, so the effect of rotation is that
$|b_{\mathrm{min}}| > |b_{\mathrm{max}}|$. As a result, rotation
allows an observer to see more of the side of the star which is moving
away from the observer, as shown in Figure~\ref{fig:angles}. In this
figure, $b_S$ corresponds to the maximum value of the impact parameter
allowed for a static star.

\subsection{Deflection of Photons}
\label{sec:defl}
\begin{figure*}
\epsscale{0.5}
\plotone{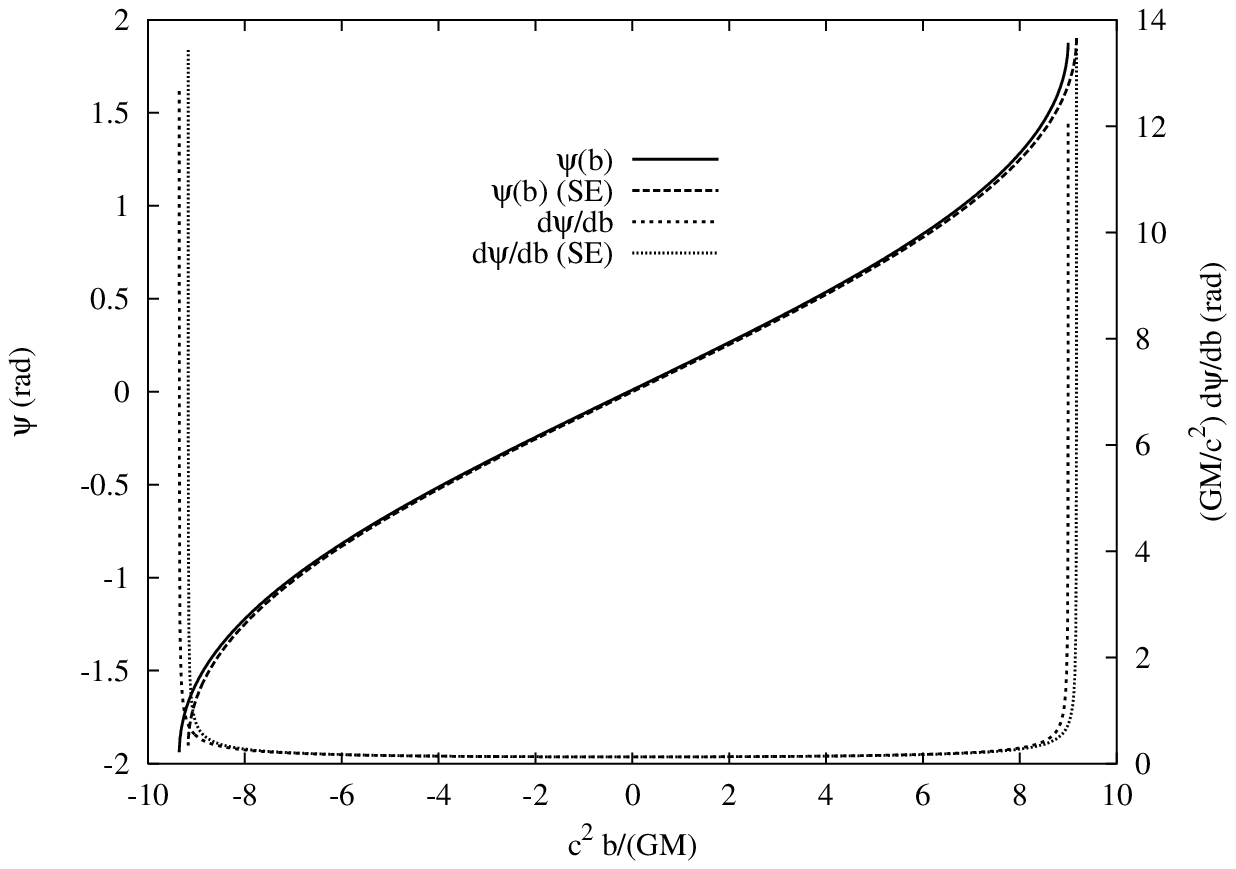}
\caption{Bending angle and its derivative as a function of impact parameter for Model 4, calculated using the full metric and the SE metric.}
\label{fig:bend}
\end{figure*}

In Figure~\ref{fig:angles} we illustrate the deflection of photons
from the point of emission on the star to the observer.  We define the
azimuthal location of the distant observer to be at $\phi = 0$. A
photon with impact parameter $b$ hits the observer if it was emitted
at azimuthal angle $\phi_i$.  The initial emission location is found
by dividing Equation~\ref{eq:nullphi} by Equation~\ref{eq:nullr}, and
integrating from the star's surface to the distant observer:
\be
\label{eq:defl}
-\phi_i(b) = \int_{\bar{r}_{e}}^{\infty} e^{\alpha-\frac12(\gamma+\rho)}
			\frac{\omega (1 - \omega b) + b e^{2\rho}/\bar{r}^{2}}
				{ \left( (1-\omega b)^{2} 
					- b^{2}  e^{2\rho}/\rb^{2} \right)^{1/2}}
			\,\ud \bar{r}.
\ee
The deflection angle $\psi$ is defined by $\psi = -\phi_i$. In the
calculation of flux from a star, both the quantities $\psi(b)$ and
$\ud\psi/\ud b$ are of importance. These quantities are plotted for
the fiducial stellar model in Figure~\ref{fig:bend}.  In addition, we
show the deflections for the SE static model. The differences between
the calculations with and without rotation are very small. The worst
errors occur at the limbs of the star, so these differences are only
likely to be of importance if the light is preferentially emitted in
directions close to the horizontal.

\subsection{Times of Arrival}
\label{sec:toa}
To accurately model pulse shapes, we account for the different amounts
of coordinate time that photons emitted from different regions of the
star will take to reach the observer.  Once the times of arrival (TOA)
are known, the photons can be placed into the correct detector timing
bins. The choice of zero time is arbitrary, so we have chosen a value
of zero TOA for a photon with zero impact parameter.  For photons
emitted with the maximal values of impact parameter, the TOA is
similar to the light travel time across the star.  For our fiducial
stellar model with $R = 16.38$ km, the light travel time is close to 80
$\mu$s. Compared to a spin period of 1.6 ms, this corresponds to a 5\%
effect, which will be seen to have a significant effect on the
calculated pulse shapes.

\begin{figure*}
\epsscale{0.5}
\plotone{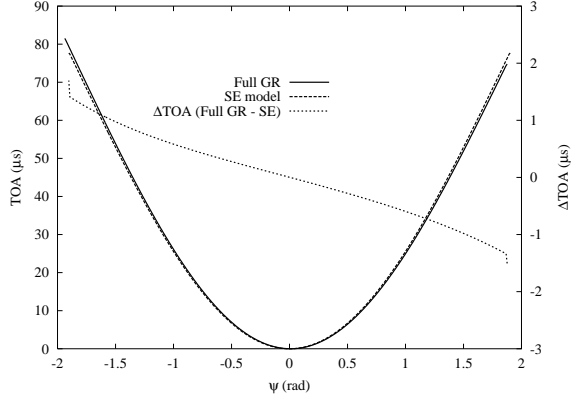}
\caption{Times of arrival (TOA) as a function of bending angle for Model 4, calculated using the full metric and the SE metric.}
\label{fig:toa}
\end{figure*}

The TOA is calculated by dividing Equation~\ref{eq:nullt} by
Equation~\ref{eq:nullr}, integrating from the star's surface to the
distant observer and then subtracting off the corresponding quantity
for a $b=0$ photon.  This yields the TOA formula
\be
\mathrm{TOA}(b) = \int_{\bar{r}_e} ^{\infty} e^{\alpha-\frac12(\gamma + \rho)}
\left( \frac{(1 - \omega b) \rb}		
     {\left( (1-\omega b)^{2} \rb^2  - b^{2}e^{2\rho} \right)^{1/2}}
			- 1 \right)\,\ud \bar{r}.
\label{eq:toa}
\ee
In Figure~\ref{fig:toa} we plot the TOA for our fiducial star and the
SE static star. Note that in the full rotational calculation, the
retrograde photon takes longer to reach the observer than the prograde
photon. This is due to the frame-dragging effect. The magnitude of
this effect is about $1/10$ of the effect due to the light-crossing
time in the corresponding SE models so we expect that for most timing
applications that it will not be detectable.

\subsection{Redshift}
\label{sec:red}
In our units, the photon's energy as measured by an observer far from
the star has been normalized to unity. As a result, any other observer
with four-velocity $u^a$ measures a photon energy of $E_u = -\ell^a
u_a$, where the photon's four-velocity components $\ell^a = \ud
x^a/\ud \lambda$ are given in Equations \ref{eq:nullt}--\ref{eq:nullr}. 
The star's four-velocity at the equator is
\be
u^a = \frac{1}{V} \left( t^a + \Omega_\star \phi^a\right),
\label{eq:fluidvel}
\ee
where $\Omega_\star$ is the star's angular velocity as measured by an
observer at infinity, and the normalization condition $u^au_a=-1$
yields
\be
V^2 = e^{\gamma +\rho} \left( 1 - (\Omega_\star-\omega)^2 \rb^2 e^{-2\rho} \right),
\ee
where all quantities are evaluated on the star's equator.

The redshift factor $(1+z)$ between light emitted at the star's
equator and detected by an observer at infinity is
\be
1+z = e^{-\frac12(\gamma +\rho)} \frac{ (1 - \Omega_\star b)}
{
\sqrt{ 1 - (\Omega_\star-\omega)^2 \rb^2 e^{-2\rho} }
}. 
\label{eq:red}
\ee
Note that the quantity 
$v_{\mathrm{ZAMO}}^{2} \equiv (\Omega_\star-\omega)^2 \rb^2 e^{-2\rho}$ 
appearing in the denominator is the square of the velocity of the star's
fluid as measured by an observer with zero angular momentum; i.e.,
an observer with  $u_\phi = u^r = u^\theta = 0$.

\subsection{Zenith Angle} 
\label{sec:zenith}
If the radiation is not emitted isotropically, the intensity will
depend on the initial direction of the photon with respect to the
fluid. The angle between the local normal to the star's surface and
the direction in which a photon is emitted is the zenith angle and
will be denoted $\alpha$.  This angle depends on the frame in which it
is measured.  For example, consider the co-moving observer at the
equator with four-velocity given by Equation~\ref{eq:fluidvel}.  In
order to define this angle in an invariant manner, we first define the
spatial projection operator $h_{ab}$ for the co-moving observer as:
\be
h_{ab} \equiv g_{ab} + u_a u_b.
\ee
Any photon's four velocity $\ell^a$ can then be decomposed into a
``time'' component $\ell^u = -\ell^a u_a = (1+z)$ and a ``spatial''
component $\ell_{\perp}^a = \ell^b h^a_{\;b}$ using the spatial
metric.  For all photons, the spatial components satisfy the
normalization condition $|\ell_\perp| = \ell^u$.

The unit normal to the fluid $n^a$ (in the equatorial plane) as
measured by the co-moving observer has only a radial component 
$n^r = e^{-\alpha}$. The co-moving observer measures an angle of $\alpha_e$
between the photon and the normal, defined by the equation
\ba
\cos \alpha_e &=& \frac{h_{ab} \ell_\perp^a n^b }{\ell^u} \nonumber\\
&=& \left( (1-\omega b)^{2}-\frac{b^{2}}{\bar{r}^{2}} e^{2\rho} \right)^{1/2}
 \frac
{\sqrt{ 1 - (\Omega_\star-\omega)^2 \rb^2 e^{-2\rho} }}{( 1 - \Omega_\star b)}.
\label{eq:comalpha}
\ea
It will also be useful for us to know what the non-rotating observer with
$u^a \propto t^a$ at the surface of the star measures for the angle formed
by the photon and the normal; following the same procedure, this angle is
\be
\cos \alpha_{o} 
= \left( (1-\omega b)^{2}-\frac{b^{2}}{\bar{r}^{2}} e^{2\rho} \right)^{1/2} 
	\sqrt{ 1 - \omega^2 \rb^2 e^{-2\rho} }.
\label{eq:obsalpha}
\ee

\subsection{Angles between Photons}
\label{sec:angles}
In a more general calculation of flux from a two-dimensional emitting
area on the star, we would need to calculate the solid angle subtended
by the area, as viewed by the observer at infinity.  In this paper, we
are only including the flux of photons emitted from a segment of the
equator into the equatorial plane.  Adopting this special
one-dimensional emission region means that observed radiation will
subtend zero solid angle in the observer's sky.  The most
straightforward way of adjusting the usual definition of flux for this
simplified emission region is to define flux in terms of an integral
over angle in the observer's one-dimensional ``sky'' which coincides
with the equatorial plane.

In Figure~\ref{fig:angles} we show a curve of angular extent
$\Delta\phi$ on the star, and the angle measured by the observer at
infinity between the two photons emitted from the end-points (points
$\phi_{-}$ and $\phi_{+}$) is $\ud\varepsilon$. If the impact parameters
for these two photons are related by $b_{+} = b_{-} + \ud b$, the angle
observed between the two photons reduces at infinity to
\be
\ud\varepsilon = \frac{\ud b}{r},
\label{eq:angles}
\ee
if both photons are restricted to move only in the equatorial plane.
This result is familiar from the usual Schwarzschild metric; its
validity for the current case is demonstrated in the Appendix.

\subsection{Outline of Numerical Method}
\label{sec:numerical}
We now outline the procedures used to compute pulse shapes from a
rotating neutron star when the photons and the observer are
restricted to the equatorial plane.  We discretise the period of the
star's rotation into $N_t$ bins and keep track of bins of flux $F_i$,
and bins of detector integration time $T_i$, where $i$ indicates the
period bin in which the flux is received.  The angular size of the
emission region $\Delta \phi$ is related to the number of period bins
by $\Delta \phi = 2\pi/N_t$.  The centre of the emission region at
each step is $\phi_i = (i-1)(\Delta \phi)$.  Figure~\ref{fig:angles}
shows the relevant quantities.

We obtain the binned fluxes $F_i$ by performing the following
steps at each period step $i$:
\begin{enumerate}
\item  Calculate the impact parameters of the null rays arriving
	at the observer from $\phi_- = \phi_i - (\Delta \phi)/2$, $\phi_i$,
	and $\phi_+ = \phi_i + (\Delta \phi)/2$.  Denote these impact
	parameters by $b_{-}$, $b$, and $b_{+}$.  This is done
	by numerically solving Equation~\ref{eq:defl}.
\item Calculate the redshift $z(b)$ using  Equation~\ref{eq:red}.
\item Calculate the zenith angle $\alpha_{e}(b)$ using Equation~\ref{eq:comalpha}.
\item Calculate the angular contribution to the flux integral,
	$\ud \varepsilon$, using Equation~\ref{eq:angles}.
\item Calculate the TOAs for the limbs,
	$\mathrm{TOA}(b_-)$ and $\mathrm{TOA}(b_+)$, expressed in 
	units of the rotation period of the star, and
	$\Delta T = |\mathrm{TOA}(b_-) - \mathrm{TOA}(b_+)|$ .  
	This is done via Equation~\ref{eq:toa}.
\item Convert the TOAs of the limbs to indices of the flux
	binning.  This is done, schematically, by computing
	$j_\pm = \mathrm{floor}(i + (N_t)(\mathrm{TOA}(b_\pm)))$, 
	where $j_\pm$ is the bin corresponding to the arrival time
	of the flux corresponding to $b_\pm$.
\item If $j_+ \neq j_-$, the flux from the limbs arrives during
	different period bins, and one needs to calculate
	the fractional amount of flux to be assigned to each
	bin, $x_{\pm}$, with $x_+ + x_- = 1$.
\item Calculate the flux integral 
	\be
	F_{\mathrm{(1-D)}} = \int_{(1+z) \nu_{o_{\mathrm{low}}}}^{(1+z) \nu_{o_{\mathrm{high}}}}
		\ud \nu_{\mathrm{e}} \int \ud\varepsilon\,I_{\nu_e}(\alpha_e) / (1 +z(b))^{4},
	\ee
	where $\nu_{o_{\mathrm{low}}}$ and $\nu_{o_{\mathrm{high}}}$
	correspond to the lower and upper limits of the detector's
	energy band.  Note that this integral is a non-standard
	definition of flux which we must adopt because we are dealing
	with a special one-dimensional emission region, as discussed
	in Section~\ref{sec:angles}; the observer only receives
	photons from within the equatorial plane (the sky is
	effectively one-dimensional), so the integral is over a
	one-dimensional angle, not the usual solid angle for a
	two-dimensional sky.  This integral simplifies in the case of
	bolometric flux since the integral will be over all energies
	at each step.  Since the angle $\ud\varepsilon$ corresponds to
	an angle between photons arriving at different times, this
	corresponds to the energy deposited in the detector in the
	time period $\Delta T$. This flux is then portioned into the
	flux bins $F_{j_\pm}$ using $F$ and $x_{\pm}$.
\item The bins of integration time $T_{j_{\pm}}$ are incremented by $(\Delta T) x_{\pm}$.
\end{enumerate}
Once the calculation for each period step is completed, the non-zero
bins of flux $F_i$ are weighted by dividing by the corresponding
integration time $T_i$.  This step is necessary because without this
weighting the fluxes in each bin do not necessarily correspond to
uniform intervals of integration time by the detector.

The fluxes $F_i$ correspond to the pulse shape for an infinitesimal
spot of angular size $\Delta \phi$.  To investigate the pulse shapes
of wider spots of width $k (\Delta \phi)$ we can compute the new pulse 
shape $P_i$ by
\be
P_i = \sum_{j=0}^{k} F_{(i-k/2) + j},
\ee
where the values $F_i$ are understood to be periodic in the index $i$.
Choosing the offset $(i-k/2)$ above is arbitrary; this choice keeps 
the spot initially centred as far as possible on $\phi = 0$.

The \texttt{rns} code calculates the metric potentials to a finite
value of $\rb$, and so our calculation is performed at distant 
$r \approx 10^{10}$~cm and not at infinity.  We have fully accounted for
the very small corrections to the expressions appearing in
Section~\ref{s:rapid} which result from not locating our observer at
infinity.

\begin{figure*}
\epsscale{0.5}
\plotone{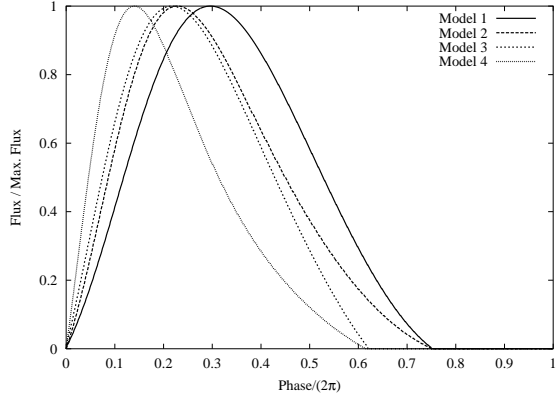}
\caption{Bolometric pulse shapes for the four model stars.  The beginning of the rise for each pulse has been aligned with Phase = 0.}
\label{fig:bolshapes}
\end{figure*}

In Figure~\ref{fig:bolshapes} the bolometric pulse shapes computed
using this algorithm are shown for the four model stars. In these
calculations, the spot size is 0.25\textdegree\ for the 300~Hz models,
and 0.5\textdegree\ for the 600~Hz models; these choices were made so
that the bin widths for these cases corresponded to equal intervals of
time.  The general effect of rotation is to create an asymmetry in the
pulse shape that increases as the rotational velocity increases.  We
have normalized all flux to the peak flux, and the moment the emission
region becomes visible is defined to be the start of the period.  We
have broken each pulse into four time periods: the rise time, the fall
time, the total time on and the total time off. These periods are listed in
Table~\ref{tab:boltimes}.

\begin{deluxetable}{crrrrrr}
\tablecaption{Timing of Bolometric Pulse Shapes
			\label{tab:boltimes}}
\tablehead{
\colhead{Model} &
\colhead{Period} &
\colhead{Rise} &
\colhead{Fall} &
\colhead{On} &
\colhead{Off}&
\colhead{$t_{\mathrm{rise}}/t_{\mathrm{fall}}$}
}
\startdata
1 & 3333.3 & 983.8 & 1511.6 & 2495.4 & 838.0 & 0.65\\
2 & 1666.7 & 377.3 & 868.1 & 1245.1 & 421.3 &  0.43\\
3 & 3333.3 & 719.9 & 1342.6 & 2062.5 & 1270.8 &0.54\\
4 & 1666.7 & 231.5 & 787.0 & 1018.5 & 648.1 &  0.29\\
\enddata
\tablecomments{All times are in $\mu\mathrm{s}$,
		and pulse shape parameters are given to $\pm 2.3\,\mu\mathrm{s}$, corresponding
		to the width of one period bin.}
\end{deluxetable}

In Table~\ref{tab:boltimes} it can be seen that increasing the star's
compactness $M/R$ increases the fraction of time that the pulse is
on. This is due to the gravitational bending of light increasing the
fraction of the time that the spot is visible. This effect is almost
independent of the star's spin rate. The asymmetry between the rise
and fall time of the pulse is shown by the column
$t_{\mathrm{rise}}/t_{\mathrm{fall}}$ in Table~\ref{tab:boltimes}. The
rise time corresponds to time during which the spot first appears on
the blueshifted limb to the time of peak intensity.  The Doppler
boosting effect shifts the phase of maximum intensity to the
blue-shifted side of the star, so that the rise time takes less than
half of the total time on. Increasing the speed increases the
asymmetry, as can be seen by comparing Tables~\ref{tab:modelparams}
and \ref{tab:boltimes}.

\section{Approximate Methods for Pulse Shape Calculations}
\label{s:approx}
Approximate methods for pulse shape calculations involve approximating
the spacetime containing the neutron star by a metric which is
analytically known.  For example, in the S+D approximation,
light-bending and time-delays are calculated as though the star was
not rotating, using the method described in \citet{PFC83}.  Special
relativistic effects are included by modeling the star as a rapidly-rotating 
object with no gravitational field.  The photons' redshifts
are computed by combining the Schwarzschild gravitational redshift
with the special relativistic Doppler effect. This approximate
approach is attractive since the gravitational field is completely
described by the Schwarzschild metric, for which simple formulae
exist. The S+D approximation has been described in detail by
\citet{PG03}, so only a brief outline of the method will be given
here. However, it should be noted that \citet{PG03} do not include
time-delays due to different light travel times from different parts
of the star.  In Section~\ref{sec:sd}, we show how the method
employed by \citet{PG03} can be recovered from the more general case
of Section~\ref{s:rapid} by taking appropriate limits.

An alternative, but similar approximation is the use of the Kerr black
hole metric to approximate the gravitational field of a rotating
neutron star \citep{CS89,BRR00,Bhat04}.  While the gravitational
fields outside of rotating neutron stars and black holes are not the
same for arbitrarily fast rotation, the Kerr black hole is a
reasonable approximation to a neutron star if it is spinning
slowly. We expect that the Kerr metric is a reasonable approximation
for stars with spin frequencies of 300 Hz, but at 600 Hz there would
be large differences.  These differences will be quantified elsewhere.

\subsection{The Schwarzschild + Doppler Approximation}
\label{sec:sd}
The S+D approximation seeks to treat the exterior spacetime of the
neutron star as though it is described by the corresponding SE
spacetime.  Effects due to the relative velocity of the star and
observer are put in by hand.  Consider the SE spacetime described by
the potentials $\alpha$, $\gamma$ and $\rho$, with $\omega = 0$,
where the relationship of the potentials to the star's mass and radius
is given in Equations~\ref{eq:lim1}--\ref{eq:lim3}.  We imagine that
the matter comprising the star is still rotating, but take the
Schwarzschild metric to be an approximation of the exterior spacetime.
Denote the Schwarzschild redshift as $z_s$ so that
\be
1 + z_s = \frac{1}{\sqrt{1 - 2M/R}}.
\ee
Using Equation~\ref{eq:lim1}, we see that the factor of
$\exp{-(\gamma+\rho)/2}$ reduces to $1+z_s$.  The equatorial velocity
of the star measured in the SE spacetime of Equation~\ref{eq:veq} can
be written, via Equation~\ref{eq:lim2}, as 
$v_{eq} = \Omega_\star \rb \exp(-\rho)$.  Introducing the function
\be
\eta \equiv \frac {\sqrt{1-v_{eq}^2}}{1-\Omega_\star b},
\label{eq:etadef}
\ee
we see that the Schwarzschild case ($\omega\rightarrow 0$) 
of the exact redshift expression in Equation~\ref{eq:red}
is given by
\be
\lim_{\omega\rightarrow0} (1+z) =  \frac{(1+z_s)}{\eta},
\label{eq:sered}
\ee
so that $\eta$ plays a role similar to the Doppler boost factor 
in special relativity with no gravitational field present.

Now we express $\eta$ in terms of the zenith angle of
Section~\ref{sec:zenith}.  As measured by the co-moving observer and
the static observer, these angles are respectively
\be
\lim_{\omega\rightarrow0}\cos \alpha_e = \eta \sqrt{1-\frac{b^2}{R^2}\left(1-\frac{2M}{R}\right)},
\label{eq:alpe}
\ee
and
\be
\lim_{\omega\rightarrow0}\cos \alpha_{o} = \frac{1}{\eta} \lim_{\omega\rightarrow0}\cos \alpha_e 
	= \sqrt{1-\frac{b^2}{R^2}\left(1-\frac{2M}{R}\right)}.
\label{eq:alpo}
\ee
Equation~\ref{eq:alpo} is a quadratic equation for $b$ in terms of
$\alpha_o$ with the solution
\be
b = \pm \frac{R}{\sqrt{1-2M/R}} \sin \alpha_{o},
\label{eq:solb}
\ee
where the plus/minus sign corresponds to co- and counter-rotating
photons.  If we know the angle $\alpha_{o}$ for a given photon, and
solve for the corresponding $b$ via Equation~\ref{eq:solb}, then using
the definition of $\eta$ from Equation~\ref{eq:etadef} and our expression for
velocity in the SE spacetime from Equation~\ref{eq:veq}, we have that
\be
\eta = \frac{\sqrt{1 - v_{eq}^2}}{1 \mp v_{eq} \sin \alpha_{o}}.
\ee
This is the usual Doppler factor for redshift without a gravitational field
present.  This formula, as well as the relationship between $\alpha_e$
and $\alpha_o$ in Equation~\ref{eq:alpo}, agree with the corresponding
formulae used by \citet{PG03} once it is noticed that they
respectively denote our angles $\alpha_e$ and $\alpha_o$ by $\alpha'$
and $\alpha$, and that the four-velocity of the fluid is perpendicular
to the normal in the observer's frame.  The overall redshift,
Equation~\ref{eq:sered}, is just the product of the Doppler factor
arising from emission in the co-moving frame into the observer's frame
at the surface, and the Schwarzschild gravitational contribution to
redshift as the photons travel to infinity.

Finally, we turn our attention to the matter of time-delays.  One is
left with the choice of accounting for the contribution of time delays
according to the method described in Section~\ref{sec:numerical}, or
else simply calculating flux directly as a function of the emitting
region's location.  If we elect to do the latter, we can still account
for the special relativistic ``snapshot'' effect; this effect arises
because the image in the detector at a given instant is formed by
photons emitted at different times from different parts of the star.
These corrections are put in by hand, making use of the conformality
of the Lorentz transformations discussed by \citet{Ter59}. The main
effect of special relativity on an area element of the star is that
the observed shape is not changed, but the element will be magnified
if it is moving away from the observer.  As discussed by \citet{PG03},
the net result is that in the S+D approximation, the solid angle must
be calculated using the formula
\be
\ud\Omega_{S+D} = \eta\,\ud\Omega_{o} = \eta\, b \, \ud b\, \ud\phi/r^2,
\ee
where $\ud\Omega_{o}$ is the uncorrected element of solid angle.  Note
that such a correction need not be applied to the solid angle element
when the flux is calculated using the method of
Section~\ref{sec:numerical} as this method accounts for all effects
due to time-delay, i.e., the snapshot effect and the light-crossing
time, by correctly binning the flux emitted by a region of constant
angular size.  To be clear, when we say a calculation ``includes
time-delay effects,'' we mean in the sense of Section~\ref{sec:numerical};
calculations that do not include these time-delay effects have
the special relativistic Doppler correction applied instead.

Since our emitting area is actually one-dimensional, we need to derive
the analogue of the S+D solid angle element for one
dimension.  The calculation for one dimension yields the same
Doppler boost factor,
\be
\ud\varepsilon_{S+D} = \eta \,  \ud b/r = \eta \frac{\ud\psi}{r\,\ud\psi/\ud b}, 
\ee  
where $\ud\psi/\ud b$ should be evaluated using the Schwarzschild
metric and the angle $\ud\psi$ is the constant angular size of the
spot.

\subsection{Comparison of Methods}
\label{sec:compare}
\begin{figure*}
\epsscale{0.5}
\plotone{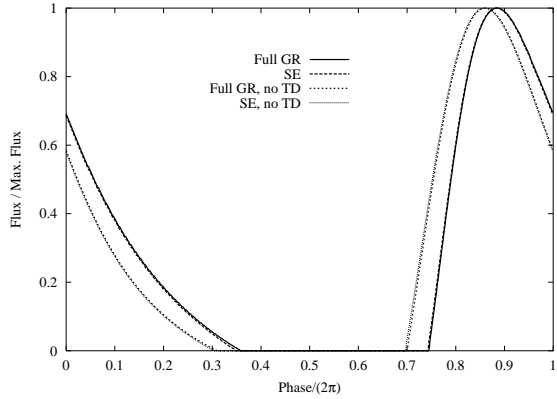}
\caption{Bolometric pulse shapes for Model 4 using different calculation methods.}
\label{fig:compare}
\end{figure*}

We have computed pulse shapes in the S+D approximation for our
fiducial star, Model~4. All calculations are for bolometric flux and
the same angular emission sizes as in Section~\ref{sec:numerical}. We
show pulse shapes in Figure~\ref{fig:compare} for four different
calculation methods: Method~I is the full general relativistic
calculation described in Section~\ref{s:rapid}. Method~II is similar
to Method~I, but the time-delay binning is not performed and the
special relativistic correction for the snapshot effect is included
instead.  Method~III is the S+D approximation including time-delay
effects.  Method~IV is the S+D approximation without time-delay
effects.

In Table~\ref{tab:compare} we show the rise, fall, on and off times
for the pulses computed with these methods. Typically, the differences
between the full GR calculation and the S+D approximation is very
small, amounting to only a few $\mu$s.  The differences between
methods that include or do not include time-delay effects is about an
order of magnitude larger.  The inclusion of time-delays keeps the
total on or off time the same within a few $\mu$s, but decreases the
rise time by 50 $\mu$s and increases the fall time by a similar
amount. This corresponds to a change of about 20\% in the rise time,
which is a resolvable amount.

\begin{deluxetable*}{llllll}
\tablecaption{Bolometric Pulse Shapes Using Different Calculation Methods
		\label{tab:compare}}
\tablehead{
\colhead{Method} &
\colhead{Rise} & 
\colhead{Fall} &
\colhead{On} &
\colhead{Off} &
\colhead{$t_{\mathrm{rise}}/t_{\mathrm{fall}}$}
}
\startdata
I   (Full GR, with TD) & 231.5 & 787.0 & 1018.5 & 648.1 &   0.29\\
II  (Full GR, without TD) & 263.9 & 747.7 & 1011.6 & 655.1 &0.35 \\
III (SE Model, with TD) & 233.8 & 775.5 & 1009.3 & 657.4 &  0.30\\
IV  (SE Model, without TD) & 268.5 & 740.7 & 1009.3 & 657.4&0.36\\
\enddata
\tablecomments{Comparison of timing parameters describing bolometric pulse shapes 
		from different calculation methods,
		using Model 4 (Period = 1.6667 ms).  Times are given to
		$\pm 2.3\,\mu\mathrm{s}$, corresponding to the width of
		one period bin.}
\end{deluxetable*}

\begin{deluxetable*}{lcc}
\tablecaption{Blackbody Soft-Hard Phase Lag Using Different Calculation Methods
		 \label{tab:bbphase}}
\tablehead{
\colhead{Method} &
\colhead{Phase Lag ($\pm 1/720$ (phase/$2\pi$))} &
\colhead{Time Lag ($\pm 2.3\,\mu\mathrm{s}$)}
}
\startdata
I (Full GR, with TD) & 0.020833 & 34.7 \\
II (Full GR, without TD) & 0.020833 & 34.7 \\
III (SE Model, with TD) & 0.020833 & 34.7 \\
IV (SE Model, without TD) & 0.019444 & 32.4 \\
\enddata
\tablecomments{ Comparison of soft/hard phase lag for unbeamed blackbody
		emission for Model 4 (Period = 1.6667 ms) for different
		calculation methods.  Radiation temperature $kT$ = 1 keV; soft band 
		is 3--5 keV; hard band is 5--7 keV.}
\end{deluxetable*}

The total time on is not changed much by the inclusion of time-delays
since the asymmetry caused by frame-dragging is quite small. However,
including time-delays increases the asymmetry between the rise
and fall times. This effect enters because the beginning of the pulse
corresponds to the moment when the spot is at the limb, where the 
time-delay is largest. However, the peak intensity corresponds to a moment
when the spot is relatively close to the observer, so the time delay is 
very small. This effectively delays the beginning of the pulse, but
not the time of peak intensity, increasing the rise/fall asymmetry. 
This can have important consequences if the pulse shape is fitted using a 
method that doesn't include time-delays, since all pulse asymmetry
will be incorrectly attributed to the rotation speed.

\section{Effects Due to the Emission Region}
\label{s:emission}
We now consider the sensitivity of the pulse shapes to various
properties of the emitting region. The different properties considered
are the emission spectrum, beaming, and the angular size of the
emitting region (``spot size'').  Unless otherwise indicated, our
calculations are for the fiducial star (EOS L, 600 Hz) using Method~I,
i.e., using the exact metric with time-delay effects.

\begin{figure*}
\epsscale{0.5}
\plotone{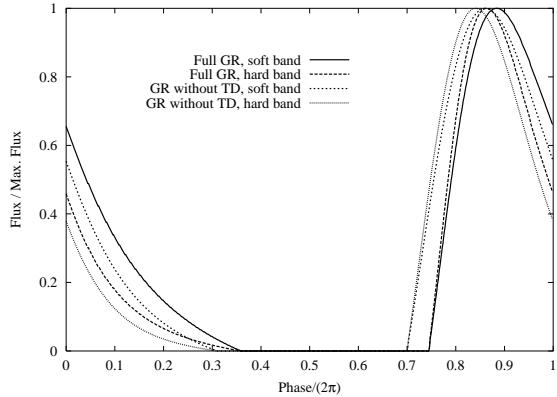}
\caption{Pulse shapes in hard and soft bands for blackbody emission using
	 different calculation methods.}
\label{fig:bands}
\end{figure*}

We computed the phase lag of soft X-rays relative to hard X-rays if
two detectors sensitive to different energy bands observe a
blackbody. We chose a blackbody of 1 keV, a soft band of 3--5 keV, and
a hard band of 5--7 keV in order to investigate this effect. In
Figure~\ref{fig:bands} we show the hard and soft pulse shapes computed
using Methods~I and II for the fiducial star; the same pulse shapes
computed using Methods~III and IV nearly overlap these curves.  The peak
of soft X-ray flux arrives after the peak of the hard X-ray flux; this
phase lag was calculated using all methods and is shown in
Table~\ref{tab:bbphase}.  The lag is independent (within error) of the
calculation method, although the individual pulse shapes do depend on
the calculation method.

\begin{figure*}
\epsscale{0.5}
\plotone{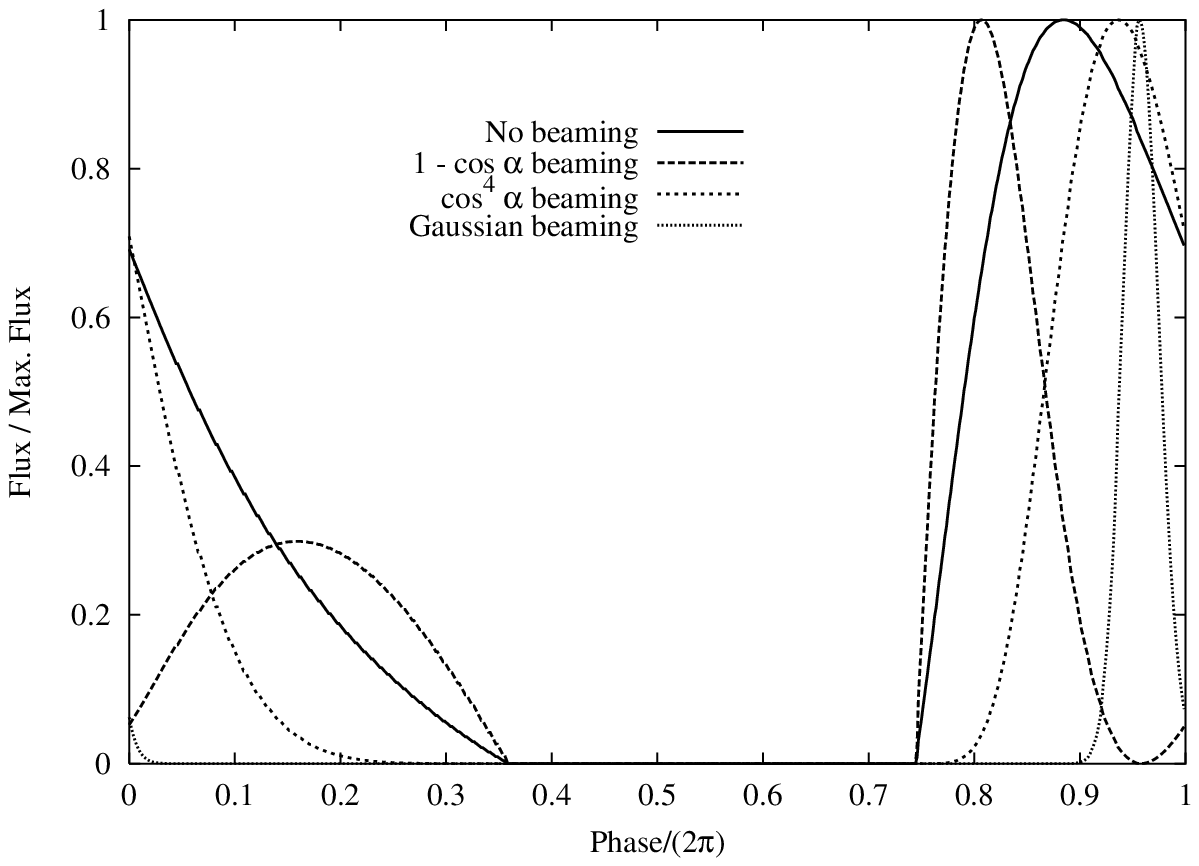}
\caption{Bolometric pulse shapes for beamed emission using Model 4.}
\label{fig:beaming}
\end{figure*}

Next we consider the effect of beaming.  We consider infinitesimal
spots which have different beaming functions and show the bolometric
flux. We show the cases of
\begin{enumerate}
\item isotropic emission; 
\item fan beaming, $I\propto 1-\cos\alpha_e$; 
\item pencil beaming, $I\propto\cos^4 \alpha_e$; and,
\item Gaussian beaming, $I \propto \exp(-\sin^2\alpha_{e}/2\sigma^2)$ 
	with $\sigma=0.1$.
\end{enumerate}
Figure~\ref{fig:beaming} shows these four pulse profiles. As the
beaming toward the normal increases, the peak intensity arrives later
since the light is emitted when the spot is closest to the
observer. Hence, an emission region composed of two components with
different types of beaming will exhibit phase lags which can be
larger than the Doppler lags in energy.

\begin{figure*}
\epsscale{0.5}
\plotone{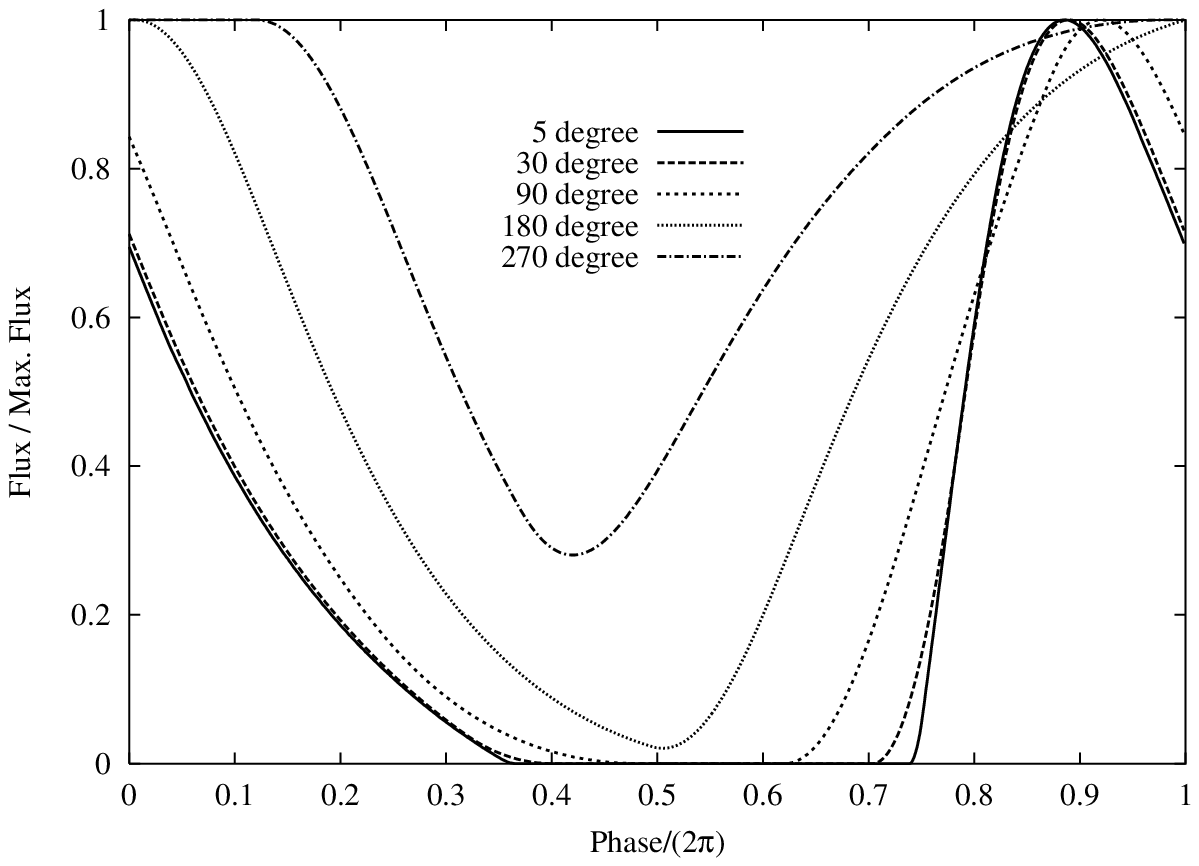}
\caption{Bolometric pulse shapes for larger emitting regions using Model 4.}
\label{fig:size}
\end{figure*}

The effect of a larger spot with uniform brightness and no beaming is
shown in Figure~\ref{fig:size}. Not surprisingly, the larger the spot size,
the broader the peak intensity.

\section{Estimation of Errors in Fitting Radius}
\label{s:fit}
We wish to estimate the error in fitting the radius of the star using
the S+D approximation.  To do this, we generated bolometric flux pulse shapes
for the following cases using Method~I:
\begin{enumerate}
\item Models 1, 2, 3, 4: isotropic emission, infinitesimal spot;
\item Models 1, 2, 3, 4: $I\propto 1-\cos\alpha_e$ emission, infinitesimal spot;
\item Models 3, 4: isotropic emission, 30\textdegree\ spot; and,
\item Models 3, 4: $I\propto 1-\cos\alpha_e$ emission, 30\textdegree\ spot.
\end{enumerate}
To these we fit pulse shapes for one-dimensional emission from an
infinitesimal region calculated using the S+D approximation without
time-delays together with either the light-bending approximation of
\citet{Bel02} or the exact light-bending formula of
Equation~\ref{eq:defl} to obtain a fitted radius. The fits performed
with the exact light-bending formula can be considered as a similar
method to that employed by \citet{PG03} for fitting observations of
\saxj\ to obtain a fitted radius; however, our method is specific to
the one-dimensional flux we defined in Section~\ref{sec:numerical} for
emission and observation restricted to the equatorial plane.  The
results are summarised in Table~\ref{tab:fitsbel} for the fits using
approximate light-bending, and Table~\ref{tab:fitsexact} for the fits
using exact light-bending.  The quality of the fits were generally
good for the cases of isotropic emission, and the fits were quite poor
for the fan beam emission.  To understand the poor quality of these
fits, note that for fan beam emission most of the flux is recorded
when the emitting region is near the limbs of the star, so the effects
due to time-delays are significant; another effect is the poor
performance of the light-bending approximation at these
locations~\citep{Bel02}, but an examination of the results in
Tables~\ref{tab:fitsbel} and \ref{tab:fitsexact} indicates that this
is a comparatively small effect on the quality of fit.  It is also
worth noting that good fits do not necessarily imply good performance
in obtaining the true radius; for example, the isotropic Model~1
calculation with approximate light-bending performs quite poorly at
obtaining the radius, while the fit was relatively good.

\begin{deluxetable*}{cclllll}
\tablecaption{ Fits to Radius Using Approximate Light-Bending Formula \label{tab:fitsbel}}
\tablehead{ 
\colhead{Method} &
\colhead{Model} &
\colhead{Emission} &
\colhead{Spot Size} &
\colhead{Fitted Radius (km)} &
\colhead{Percent Error\tablenotemark{a}} &
\colhead{$\mathrm{S.Sq.}/N$ $(\times 10^{-4})$ \tablenotemark{b}}
}
\startdata
I & 1 & isotropic                 & 0.25\textdegree & 10.517  & $+9.3$    & 0.72    \\* 
  & 2 &                           & 0.5\textdegree  & 10.344  & $+5.7$    & 0.38    \\* 
  & 3 &                           & 0.25\textdegree & 15.646  & $+3.6$    & 0.045   \\* 
  & 4 &                           & 0.5\textdegree  & 16.747  & $+2.2$    & 3.2     \\* 
II &   & 			  &		    & 16.463  & $+0.5$    & 0.14    \\* 
III &   &                           &                 & 16.788  & $+2.5$    & 2.8     \\*
IV &   &                           &                 & 16.358  & $-0.1$    & 0.25    \\[0.25em] 
I & 1 & $I\propto 1-\cos\alpha_e$ & 0.25\textdegree & 10.348  & $+7.6$    & 34      \\* 
  & 2 &                           & 0.5\textdegree  & 10.744  & $+9.8$    & 67      \\* 
  & 3 &                           & 0.25\textdegree & 15.456  & $+2.3$    & 36      \\* 
  & 4 &                           & 0.5\textdegree  & 17.765  & $+8.5$    & 59      \\[0.25em] 
  & 3 & isotropic                 & 30\textdegree   & 15.045  & $-0.4$    & 0.82    \\* 
  & 4 &                           &                 & 15.68   & $-4.3$    & 2.3     \\[0.25em] 
  & 3 & $I\propto 1-\cos\alpha_e$ & 30\textdegree   & 14.495  & $-4.1$    & 42      \\* 
  & 4 &                           &                 & 14.783  & $-9.7$    & 55      \\  
\enddata
\tablenotetext{a}{Positive values indicate fitted values are too large, negative values indicate fitted values are too small.}
\tablenotetext{b}{The sum of squared differences between fitted approximate calculation and rapid-rotation calculation divided
		  by the number of phase bins in the exact pulse shape.}
\end{deluxetable*}

\begin{deluxetable*}{cclllll}
\tablecaption{ Fits to Radius Using Exact Light-Bending Formula \label{tab:fitsexact}}
\tablehead{ 
\colhead{Method} &
\colhead{Model} &
\colhead{Emission} &
\colhead{Spot Size} &
\colhead{Fitted Radius (km)} &
\colhead{Percent Error\tablenotemark{a}} &
\colhead{$\mathrm{S.Sq.}/N$ $(\times 10^{-4})$ \tablenotemark{b}}
}
\startdata
I & 1 & isotropic                 & 0.25\textdegree & 9.70   & $+0.8$  & 0.44\\* 
  & 2 &                           & 0.5\textdegree  & 9.975  & $+2.0$  & 2.7 \\* 
  & 3 &                           & 0.25\textdegree & 15.19  & $+0.5$  & 0.50\\* 
  & 4 &                           & 0.5\textdegree  & 16.85  & $+2.9$  & 4.8 \\[0.25em] 
  & 1 & $I\propto 1-\cos\alpha_e$ & 0.25\textdegree & 9.73   & $+1.1$  & 24  \\* 
  & 2 &                           & 0.5\textdegree  & 10.28  & $+5.1$  & 56  \\* 
  & 3 &                           & 0.25\textdegree & 14.98  & $-0.8$  & 32  \\* 
  & 4 &                           & 0.5\textdegree  & 17.66  & $+7.8$  & 58  \\[0.25em] 
  & 3 & isotropic                 & 30\textdegree   & 14.44  & $-4.4$  & 0.61\\* 
  & 4 &                           &                 & 15.71  & $-4.1$  & 3.0 \\[0.25em] 
  & 3 & $I\propto 1-\cos\alpha_e$ & 30\textdegree   & 13.93  & $-7.8$  & 33  \\* 
  & 4 &                           &                 & 14.51  & $-11.4$ & 49  \\  
\enddata
\tablenotetext{a}{Positive values indicate fitted values are too large, negative values indicate fitted values are too small.}
\tablenotetext{b}{The sum of squared differences between fitted approximate calculation and rapid-rotation calculation divided
		  by the number of phase bins in the exact pulse shape.}
\end{deluxetable*}

Considering the Method~I cases with an infinitesimal spot in both sets
of fits, the radius is generally overestimated by up to 9.8\%, with
the exception of the fit to the fan-beam case for Model~3 calculated
with exact light-bending, where the result was a poor quality fit
underestimating the radius by 0.8\%.  The tendency to overestimate the
radius is a result of the fits ignoring the time-delay effect, so that
all asymmetries between the rise and fall time
($t_{\mathrm{rise}}/t_{\mathrm{fall}}$ in Tables~\ref{tab:boltimes}
and \ref{tab:compare}) are attributed to the equatorial speed of the
star.  This forces the fitting program to assume a larger speed and
therefore a larger radius (since the spin frequency is fixed).  Of the
four models computed with isotropic emission Model~4 has the worst
quality fit, which is not surprising since it was chosen to maximize
the effects of rotation.

For the 30\textdegree\ emitting region, the fits underestimate the
actual radius by up to 11.4\%.  The change in fit radius from the
infinitesimal case is $-3.8$--$-17.8$\%, which is in part cancelled by
the positive error in all but one of the infinitesimal cases.  More
compact stars have light visible for a longer fraction of their
period, and so the fit attempts to compensate for the effect of a
large emitting region (which is visible for a longer fraction of the
period) by making the star more compact.

We also performed fits with approximate light-bending to the pulse
shapes for Model 4 calculated using Methods~II, III, and IV for a
0.5\textdegree\ spot with both isotropic and fan beam emission;
Table~\ref{tab:fitsbel} shows the results for the isotropic cases.
Methods~I and II both use the exact metric, but Method~II does not
include time-delays.  The difference between the fitted radii obtained
for these methods is $0.3$~km for isotropic emission, and $1.3$~km for
fan beam emission.  Comparing the fitted radii for pulse shapes that
differ only by the selection of the metric (SE or exact), the
differences in the obtained radius ranged from $0.03$--$0.17$~km,
depending on beaming and the inclusion of time-delay effects.  In this
case the largest effect on the fitted radius comes from the inclusion
of time-delay effects, with the effect due to the metric approximation
about an order of magnitude smaller.  We expect the errors arising
from the approximation of the metric to get larger for more compact
models, i.e., softer EOS, as the frame-dragging effect can be larger
for such stars, while the time-delay effects would get smaller owing
to the shorter light-crossing time.

Finally, the error in the fitted radius obtained when we fit to the
Model 4 pulse shape calculated using the S+D approximation without
time-delays (Method~IV) was $0.02$~km for isotropic emission and
$0.1$~km for fan beam emission.  This error arises only from the
\citet{Bel02} approximation for light-bending used by the fitting
procedure.

Comparing the set of results using approximate light-bending in
Table~\ref{tab:fitsbel} to the results using exact light-bending in
Table~\ref{tab:fitsexact}, it is clear that it is not uniformly true
that fitting with the exact light-bending formula necessarily yields
improved results, improved fit quality, or both.  For example, the
calculations for Model 4 with isotropic emission and an infinitesimal
emission region is an example where the use of the exact light-bending
formula actually yields a poorer fit to the radius, and a poorer
quality of fit.

\mbox{~}

\mbox{~}

\section{Conclusions}
\label{s:conclude}
Observations of the pulse profile of light emitted from the surface of
a neutron star have the potential to constrain the star's mass,
radius, and equation of state.  Restricting ourselves to the case of
light emitted and received in the equatorial plane of a
rapidly-rotating neutron star, we have compared a variety of methods
to calculate the pulse shapes observed by a distant observer; this
restriction requires us to use a definition of flux adapted to the
one-dimensional nature of the emitting region which was employed for
all of our calculations.  In particular, we have compared the use of
an exact neutron star metric including a general relativistic
treatment of all effects due to rotation, with the use of an
associated Schwarzschild metric where the effects due to rotation of
the star are put in by hand (the ``Schwarzschild + Doppler'' (S+D)
approximation).  For each of these cases we have either included the
effects of time-delays by fully accounting for the times of arrival of
photons from different locations on the star, or else neglected the
totality of these effects and only applied a special relativistic
Doppler correction to account for the so-called ``snapshot'' effect.

The time-delay effects which we are able to account for can be split
into two classes: effects owing to the light-crossing time which are
also present for static stars, and effects owing to frame-dragging
which is present in an exact calculation of the neutron star
spacetime.  For our fiducial star, a 600~Hz, $1.4M_\odot$ neutron star
with $R=16.38~\mathrm{km}$, accounting for the light-crossing time
had the effect of decreasing the rise time and increasing the fall
time of the calculated pulse by about 50~$\mu$s, which is about 20\%
of the rise time and is a resolvable amount.  Effects due to
frame-dragging in this case were about an order of magnitude smaller.
While we found that the calculation method does change the pulse
shape, we found that the calculation of the total lag between the
peaks of flux in soft and hard X-rays for a simple blackbody emission
spectrum did not strongly depend on the calculation method used.

Constraints on the stellar parameters are obtained by fitting
theoretical pulse shapes to observations.  We wished to determine
whether a fitting routine which uses the S+D approximation without
time-delay effects could accurately determine the parameters
corresponding to pulse shapes calculated with all effects included.
Using an exact treatment of photon propagation from an infinitesimal
one-dimensional emission region, we created artificial light curves
for a number of different neutron stars and equatorial emission types.  We
then attempted to fit these light curves using a code based on the S+D
approximation without time-delays to obtain a fitted radius.
We found for infinitesimal emission regions that fitting to the
approximate pulse shapes tends to overestimate the stellar radius by
up to 9.8\%.  On the other hand, fitting to a pulse shape for an
emission region that is 30\textdegree\ wide has the effect of
underestimating the radius by up to about 10\%.  Furthermore, the
quality of these fits was quite poor in the case of emission which was
preferentially emitted in the horizontal direction.
These results should not be misconstrued as a claim that the analysis
of \saxj\ observations by \citet{PG03} is flawed since our
calculations only provide a worst-case scenario.  The X-ray pulsar
\saxj\ rotates at a slower rate than our fastest neutron star model,
and most likely emits its radiation from high latitudes where the S+D
approximation should be more valid.

In this paper we have restricted our attention to photons emitted from
the equator whose paths stay in the equatorial plane, mainly because
the effects of rotation should be strongest for these
photons. However, if the emission area is large and encompasses a wide
range of latitudes on the star, we expect that the varying shape of
the star will have an important effect. Rotation causes a star to
become oblate, and the effective gravitational field at the poles is
larger than at the equator. As a result, the static part of the
gravitational redshift is larger at the poles than at the equator. In
addition, light emitted near the poles will experience a greater
gravitationally-induced time delay. We intend to explore the effect of
rotation on pulses observed from spots and observers at a range of
latitudes in a future publication.

\acknowledgments
This research was supported by grants from NSERC. 
We also thank the Theoretical Physics Institute at the University of
Alberta for hosting the visits of DL to Edmonton.

\clearpage

\appendix
\section{Derivation of the Observed Angle Between Photons}
Equation~\ref{eq:angles} gives the infintesimal angle $\ud\varepsilon$
formed by the world lines of two photons as measured by a static
observer located in the equatorial plane at infinity.  Let the two
photons have four-velocities $\ell^a$ and $m^a$, with impact
parameters $b_+$ and $b_-$, respectively.  The angle is calculated as
in Section~\ref{sec:zenith}: The oberserver has $u^{a} \propto t^{a}$,
and we define the projection operator as $h_{ab}^{(\infty)} \equiv g_{ab} +
u_a u_b$ and the magnitudes of the projected null vectors as
$\ell^u \equiv |\ell_\perp| = |h_{ab}^{(\infty)} \ell^b |$.  The
angle is calculated by the inner product of the projected vectors:
\ba
\cos \varepsilon & = &  \frac{h_{ab}^{(\infty)} \ell_\perp^a m_\perp^b}
				{\ell^u m^u} \nonumber \\ 
		& = &
			\omega(b_+ + b_-) + b_+ b_-
			\left(\frac{e^{2\rho}}{\rb^2} - \omega^2
			\right) \nonumber \\
		&&	+ \sqrt{ \left[ (1 - \omega b_+)^{2} -
			\frac{b_{+}^2 e^{2\rho}}{\rb^2} \right] \left[
			(1 - \omega b_-)^{2} - \frac{b_{-}^2
			e^{2\rho}}{\rb^2} \right] }.
\ea
This formula is only valid when the photons are restricted to move in
the equatorial plane, i.e., $u^\theta = 0$.  To get the infinitesimal
version of this equation, put $b_+ = b_- + \ud b$, and Taylor expand
the left-hand side for small angles $\ud\varepsilon$.  Expanding the
right-hand side in $\ud b$ and equating the second-order terms gives:
\be
\ud \varepsilon^{2} = - \ud b^2
			\left[ (1 - \omega b_-)^2 
				 - \frac{ b_-^2 e^{2\rho} }{\rb^2} \right]
			\left[ \frac{ \omega^2 \rb^2 - e^{2\rho} }
				{ \rb^2\left(1-\omega b_-\right)^2 -
					b_-^2 e^{2\rho} }
				- \left(
				 	\frac{\omega^2 b_- \rb^2 - b_-
						e^{2\rho} - \omega\rb^2}
					{ \rb^2\left(1-\omega b_-\right)^2 -
						b_-^2 e^{2\rho} }
				\right)^2
			\right].
\ee
In the large $\rb$ limit, $\omega$ falls off as $1/\rb^3$, so to
leading order in $1/\rb$ the first term in square brackets is 
$\sim 1$ and the second term in square brackets is 
$\sim -e^{2\rho}/\rb^2$.  So we have for large $\rb$,
\be
\ud \varepsilon = \frac{e^{\rho} \ud b}{\rb},
\ee
and in terms of the usual Schwarzschild $r$ coordinate given by
Equation~\ref{eq:lim2}, this is
\be
\ud \varepsilon = \frac{\ud b}{r},
\ee
which is Equation~\ref{eq:angles}.

\end{document}